\documentclass[]{interact}

\usepackage{epstopdf}% To incorporate .eps illustrations using PDFLaTeX, etc.
\usepackage[caption=false]{subfig}% Support for small, `sub' figures and tables

\usepackage[numbers,sort&compress]{natbib}% Citation support using natbib.sty
\bibpunct[, ]{[}{]}{,}{n}{,}{,}% Citation support using natbib.sty
\makeatletter% @ becomes a letter
\def\NAT@def@citea{\def\@citea{\NAT@separator}}% Suppress spaces between citations using natbib.sty
\makeatother% @ becomes a symbol again

\theoremstyle{plain}% Theorem-like structures provided by amsthm.sty

\theoremstyle{definition}

\theoremstyle{remark}

\begin{document}

\articletype{Regular Article}% Specify the article type or omit as appropriate

\title{The use of LEDs as a light source for fluorescence pressure measurements}

\author{
\name{Rustem~Khasanov\textsuperscript{a}\thanks{CONTACT Rustem Khasanov. Email: rustem.khasanov@psi.ch},
    Matthias Elender\textsuperscript{a}\thanks{CONTACT Matthias Elender. Email: matthias.elender@psi.ch}, and
    Stefan Klotz\textsuperscript{b}\thanks{CONTACT Stefan Klotz. Email: stefan.klotz@upmc.fr}}
\affil{\textsuperscript{a}Laboratory for Muon Spin Spectroscopy, Paul Scherrer Institute, CH-5232 Villigen PSI, Switzerland \\ \textsuperscript{b}Sorbonne Universit\'{e}, CNRS UMR 7590, Institut de Min\'{e}ralogie, de Physique des Mat\'{e}riaux et de Cosmochimie (IMPMC), Paris, France}
}

\maketitle

\begin{abstract}
We discuss the use of commercial high-power light emitting diodes (LEDs) as a light source for fluorescence pressure measurements. A relatively broad light emitting spectra of  single color LEDs (in comparison with lasers) do not prevent producing narrow fluorescence lines at least for two widely used pressure indicator materials, namely ruby (Cr$^{3+}$:Al$_2$O$_3$) and strontium tetraborate (Sm$^{2+}$:SrB$_4$O$_7$). Strongest responses of both pressure indicators were detected for the green color LEDs with the average wavelength $\lambda_{\rm av}\sim 530$~nm. LEDs might be easily implemented for producing fiber coupled, as well as the parallel light sources. LEDs were found to be efficient to replace laser sources in piston-cylinder cell and diamond anvil cell fluorescence pressure measurement setups.
\end{abstract}


\section{Introduction}

Pressure is a universal tool to tune the electronic structure and, by doing this, to study in a clean and controllable way the coexistence and interplay of different energy scales in a broad variety of materials. In some particular cases, the external pressure may lead to formation of a new, previously unknown, structural phases with unique physical properties which were not detected at ambient pressure conditions. One of the major example is the observation of a nearly room temperature superconductivity in hydrogen rich materials, where the new phases were synthesized inside diamond anvil cells by keeping them  at hundreds of gigapascal pressures (see {\it e.g}. Refs.~\cite{Drozdov_Nature_2018, Boeri_JPCM_2022} and references therein).

One of the drawbacks of pressure experiments is  the necessity of using especially dedicated containers (cells), which are suitable for generating high pressures. The key question, in such case, becomes the knowledge of an accurate pressure value ($p$) inside the cell, which, in most of the cases, differs from the simplified assumption of $p=F_{\rm ap}/S$ (where, $F_{\rm ap}$ and $S$ are the applied force and the area to which this force is being applied, respectively). The most direct way is to look on the response of a so-called pressure indicator, which might be placed inside the cell chamber together with the sample under investigation. Such pressure indicator needs, on the one hand, to be small in order not to occupy much of the cell volume available for the sample and, on the other hand, to be sensitive to the applied pressure.

One of the widely used pressure determination techniques is based on the effect of fluorescence.\cite{Forman_Science_1972, Mao_JGR_1986, Hess_JAP_1992, Datchi_JAP_1997, Cao_SrB4O7_ApplMatt_2016, Shen_HPR_2021, Podlesnyak_HPR_2018, Naumov_PRA_2022}  In such case, the absolute pressure value in the close vicinity of the sample is determined by detecting the pressure induced shift of fluorescence line(s) of the pressure indicator (typically containing  Cr$^{3+}$ or Sm$^{2+}$ ions). The technique requires the use of an external light source, a special setup allowing to separate optical passes of the direct (excitation) and reflected (fluorecense) lights, and an optical spectrometer.
In current fluorescence systems the excitation source involves, typically, a laser source in the spectral range from the dark blue to the near red.\cite{Shen_HPR_2021, Podlesnyak_HPR_2018, Naumov_PRA_2022, Feng_RSI_2010} By supplying  intensive coherent light, the laser sources allow to measure tiny ruby chips with the dimensions not exceeding $10-20$~$\mu$m placed inside the cells.

Note that the use of laser sources should follow strict safety rules, since the laser light exposure can result in damages to the human eye and skin (see {\it e.g.} Ref~\cite{Laser-safety} and references therein). Such rules can be easily satisfied at closed laboratories, but it might be difficult to follow them at major user-based facilities among national labs. However, the coherence property of a laser, which
makes it hazardous for the eye even at very low powers, is not a critical condition for fluorescence excitation. The first pressure dependent ruby fluorescent experiments were performed by using angular dispersive light from a mercury lamp source, Ref.~\cite{Forman_Science_1972}. More recently Feng, Ref.~\cite{Feng_RSI_2011}, built an optical setup allowing to detect the response of ruby in diamond anvil cell (DAC) by using blue LED with the average wavelength $\lambda_{\rm av}\simeq 430$~nm.

In this work we studied the ability of commercial high-power LEDs to be used as a light source for the pressure determination by means of fluorescence. Our tests reveal that a relatively broad light emitting spectra of single color LEDs (compared to monochromatic laser ones) do not prevent formation of narrow fluorescence lines.  The strongest responses for two widely used pressure indicator materials, namely ruby (Cr$^{3+}$:Al$_2$O$_3$) and strontium tetraborate (Sm$^{2+}$:SrB$_4$O$_7$), are obtained for green LEDs with $\lambda_{\rm av}\sim 530$~nm. It is demonstrated that LEDs could be easily implemented for producing fiber coupled, as well as parallel collimated light sources. The latter one was found to be efficient to replace laser source in DAC setups and allow to detect a strong fluorescence response from a single ($\simeq 15$~$\mu$m in size) ruby chip.

The paper is organized as follows: Section~\ref{sec:experimental-details} describes the details of the experiment and construction of a 'fiber coupled` and 'parallel light` LED holder. Section~\ref{sec:LED-efficicncy} presents  the measurements of LEDs excitation spectra and the efficiency of fiber coupled light sources. The results of fluorescence measurements of  ruby (Cr$^{3+}$:Al$_2$O$_3$) and strontium tetraborate (Sm$^{2+}$:SrB$_4$O$_7$) pressure indicator materials are summarised in Sec.~\ref{sec:ruby-SBO_responce}.  Section~\ref{sec:replacemement-of-laser-by-LED}  discusses the possibility of replacing lasers by LED light sources in a piston-cylinder and DAC fluorescence measurement setup. The summary and conclusions follow in Sec.~\ref{sec:Conclusions}.

\section{Experimental details}\label{sec:experimental-details}

\begin{figure*}[htb]
\centering
\includegraphics[width=1.0\linewidth]{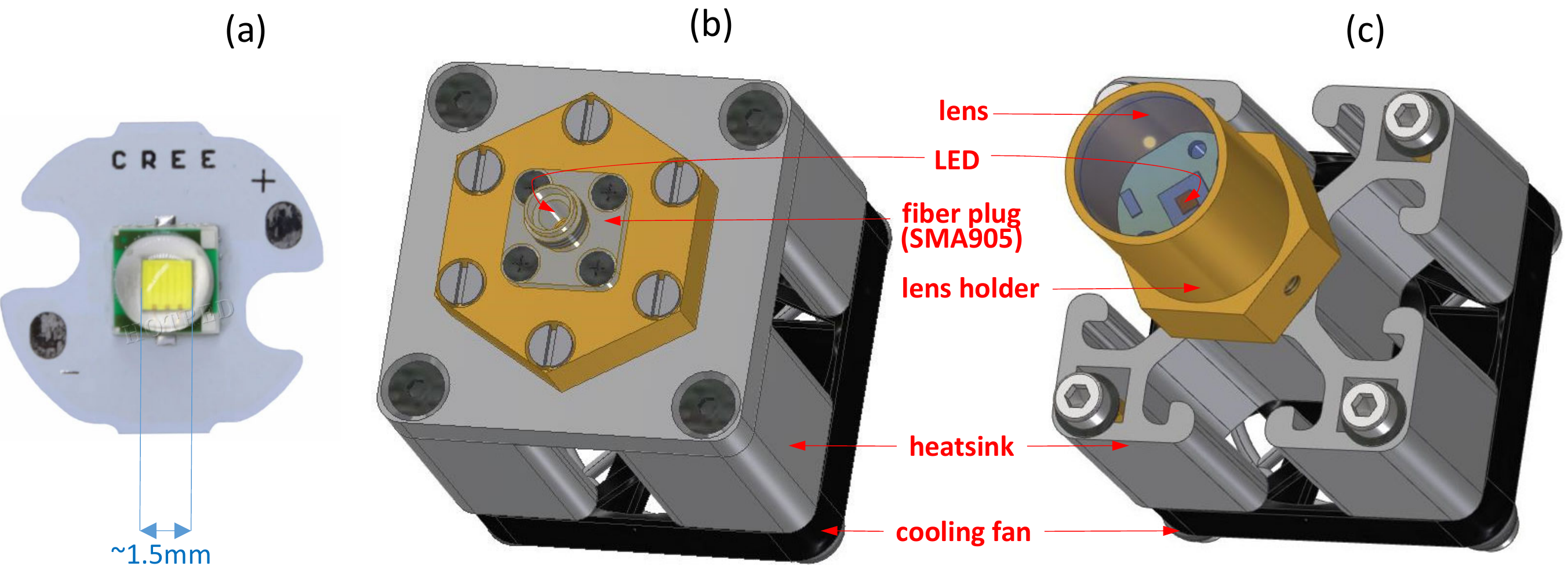}
\caption{ (a) Commercial LED mounted on a {\o}$\simeq16$~mm aluminum PCB (printed circuit board).\cite{CREE} (b) LED holder with the SMA905 optical fiber output. (c) LED holder with the 'parallel light' output. The lens focusing distance is $\simeq 18$~mm. The distance between the lens and the LED chip (from $\simeq 5$ to $\simeq 25$~mm) might be changed by replacing the lens holder. }
 \label{fig:LED_mount}
\end{figure*}

Experiments were performed by using commercial LEDs from CREE.\cite{CREE} The design with a high-power LED chip mounted on a ${\o}\simeq16$~mm aluminum PCB (printed circuit board) was selected [Fig.~\ref{fig:LED_mount}~(a)].  LEDs with the maximum electric power ($P_{\rm el}$) limited to 3~W were used for studying the efficiency of LED light to excite the fluorescence response of  Cr$^{3+}$:Al$_2$O$_3$ and Sm$^{2+}$:SrB$_4$O$_7$ and for measurements with the piston-cylinder cell. Both 3~W and 10~W LED versions were used to perform experiments with DAC.

Two type of LED mounts (holders) were constructed [Figs.~\ref{fig:LED_mount}~(b) and (c)]. Both mounts share certain parts [as, {\it e.g.}, the heat sink (made of $40\times40$~mm$^2$ aluminum profile rail), the cooling fan ($40\times40\times10$~mm$^3$, 5~V, 0.05~A), as well as several mounting pieces (screws, washers, spacers, {\it etc.})], but they are different by providing two types of optical outputs.
The holder shown in Fig.~\ref{fig:LED_mount}~(b) is equipped with a SMA905 connector and it is used for attaching optical fibers. The holder presented in Fig.~\ref{fig:LED_mount}~(c) allows creating a parallel light output. The light focusing is achieved by changing the distance between the LED and the lens via replacement the lens holder.

The optical measurements were performed by using the broad-band USB4000 (wavelength range: $350-1000$~nm)  and the narrow-band  HR2000$+$ (wavelength range: $670-770$~nm) spectrometers from Ocean Optics.\cite{OceanOptics} The optical fibers with two different types of core material (fused silica and PMMA) and various core diameters  (from 125 to 1000~$\mu$m) were probed. LEDs were powered by using a computer controlled DC6006L power supply unit from Fnirsi.\cite{Fnirsi} Experiments described in Sec.~\ref{sec:ruby-SBO_responce} were performed by using the RP22 reflection probe from Thorlabs.\cite{Reflection_probe}

\section{Results and Discussions}

\subsection{LED excitation spectra and efficiency of fiber coupled sources}\label{sec:LED-efficicncy}

\begin{figure}[htb]
\centering
\includegraphics[width=0.6\linewidth]{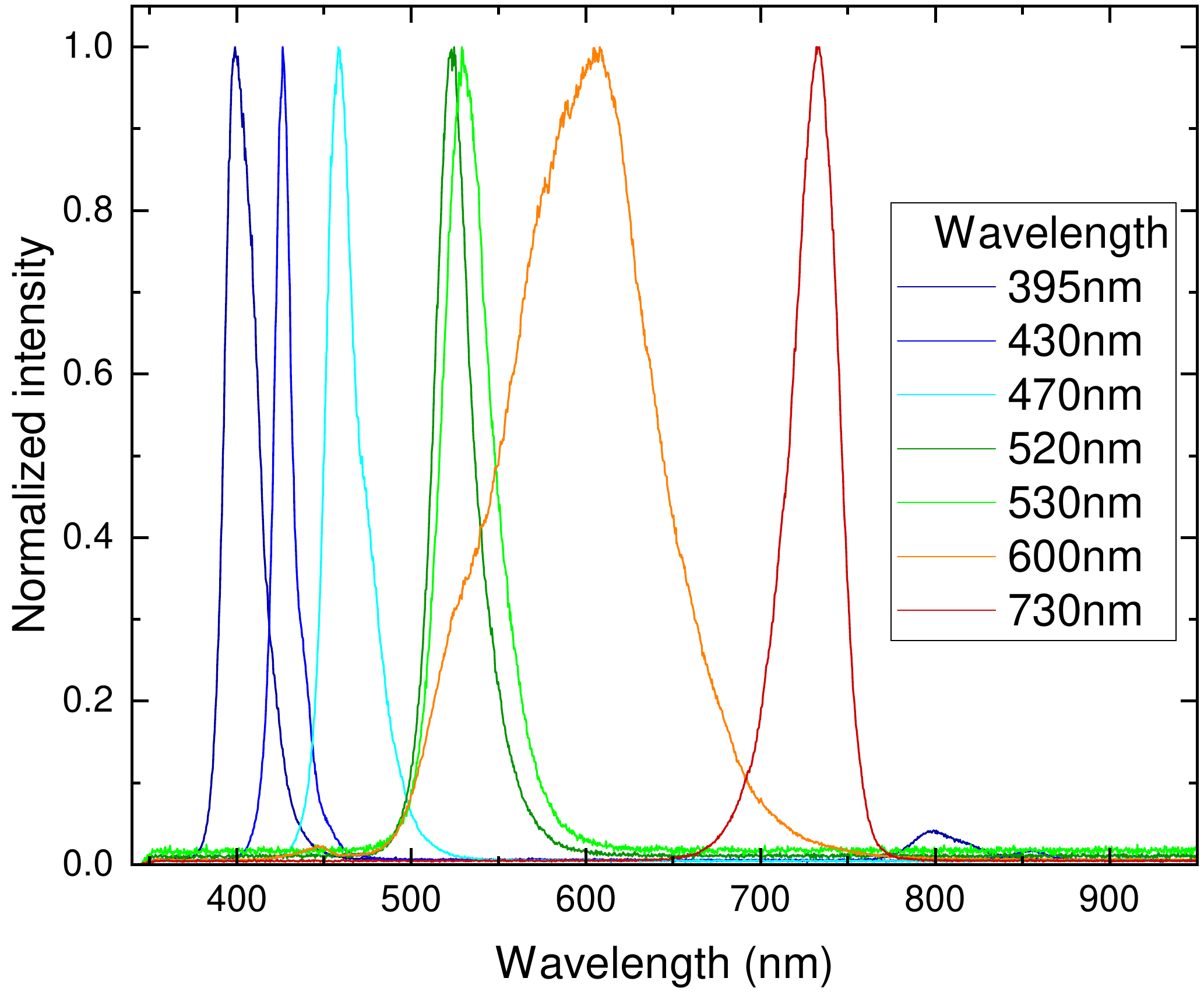}
\caption{ Excitation spectra of seven LED types used in the present study. }
 \label{fig:LED_spectra}
\end{figure}

Figure~\ref{fig:LED_spectra} shows the emission spectra of seven LED types with different wavelengths [experiments were performed by using fiber coupled LED holders, Fig.~\ref{fig:LED_mount}~(b)]. The characteristics of the studied LEDs are summarized in Table~\ref{tab1}. Nearly all LEDs (except the 600~nm one) have relatively narrow line width. The FWHM ranges from $\simeq 12$~nm (for 430~nm LED) to $\simeq 34$~nm (for 530~nm one). The exception is only the 600~nm LED with FWHM reaching $\simeq 95$~nm.

\begin{figure}[htb]
\centering
\includegraphics[width=0.6\linewidth]{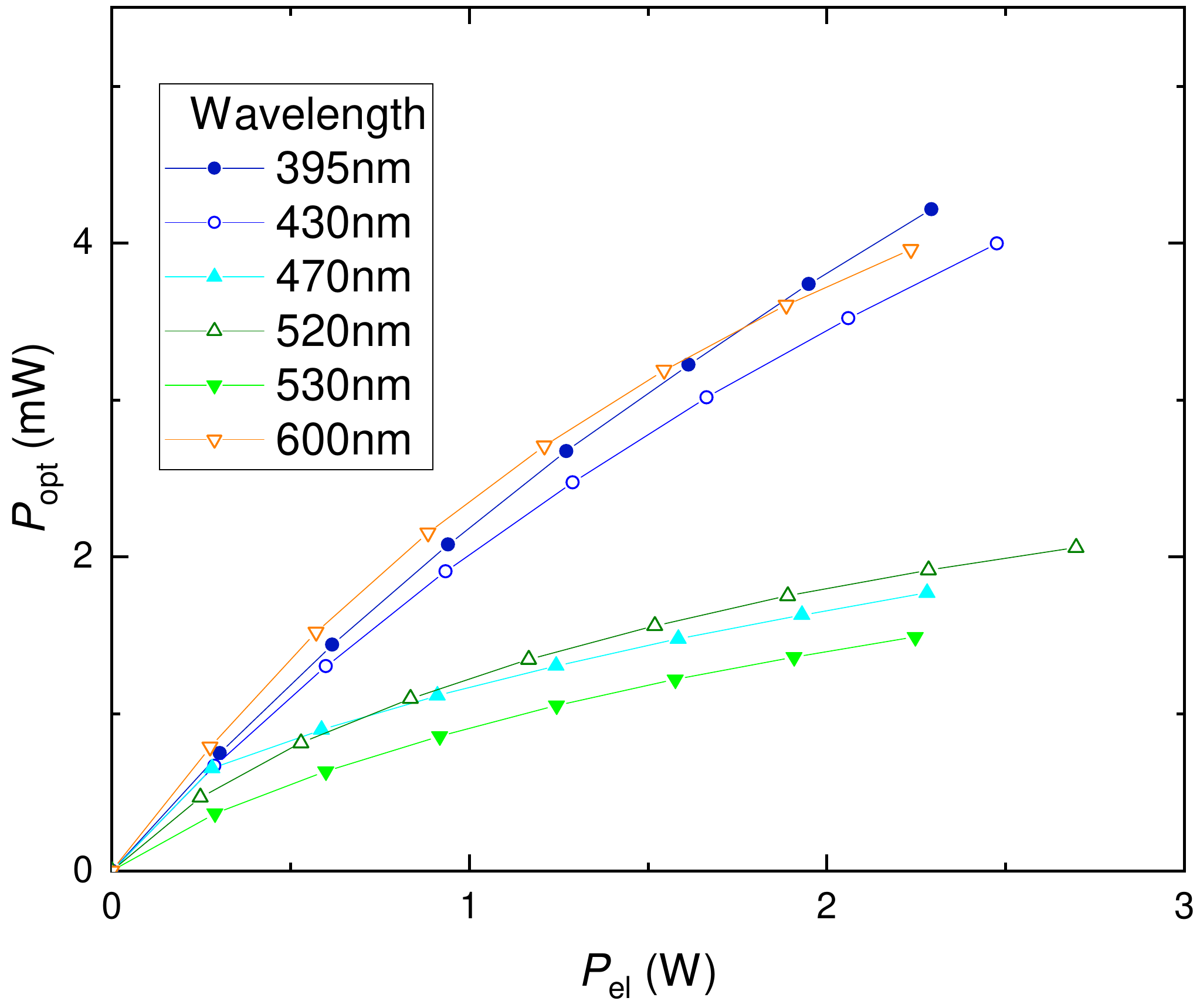}
\caption{Dependence of the output optical power ($P_{\rm opt}$) on the applied electric power ($P_{\rm el}$) for different LED types. Measurements were performed by using ${\o}=200$~$\mu$m optical fiber (the numerical aperture NA=0.29, Ref.~\cite{Fiber_200mnm}) attached to the end of the SMA905 plug [see Fig.~\ref{fig:LED_mount}~(b)]. }
 \label{fig:LED_fiber}
\end{figure}

Figure~\ref{fig:LED_fiber} shows the dependence of the output optical power ($P_{\rm opt}$) on the applied electric power ($P_{\rm el}$) for different LED types. Measurements were performed by using ${\o}=200$~$\mu$m optical fiber (the numerical aperture NA=0.29, Ref.~\cite{Fiber_200mnm}) attached to the end of the SMA905 plug [see Fig.~\ref{fig:LED_mount}~(b)].
The values of the optical power at applied current $J=0.7$~A (which corresponds to approximately  $\sim75$\% of the maximum electric power of 3~W LED) and the efficiency of fiber coupled LEDs [{\it i.e.} the ratio  $P_{\rm opt}(0.7{\rm ~A})/P_{\rm el}(0.7$~A)] are summarised in Table~\ref{tab1}.

From the above presented LED test results the following three important points emerge.
\begin{itemize}
 \item[(i)] {The full width at half maximum of studied LEDs varies between 12 and 95 nanometers (see Figure~\ref{fig:LED_spectra} and Table~\ref{tab1}). These values are similar (or  even slightly smaller) in size to the excitation bandwidths of ruby and strontium tetraborate.\cite{Mykhaylyk_ruby_Sensors_2020, Zheng_SrB4O7_JMC_2020, Cao_SrB4O7_ApplMatt_2016}  This allows to select an individual diode light source to supply the optimum excitation wavelength band for different types of pressure gauges. }

 \item[(ii)] {LEDs with the wavelengths 395, 430, and 600~nm  exhibit optical power output nearly twice as high compared to LEDs ranging from 470 to 530~nm (see Fig.~\ref{fig:LED_fiber}). This is the manifestation of the so-called ''green gap`` problem, {\it i.e.} the efficiency reduction of green LEDs compared with ultraviolet to blue and orange to infrared ones.\cite{Titkov_Materials_2017, Broell_OSRAM/LED_SPIE_2014} }

 \item[(iii)] {The area of the 3~W LED lightening chip [$\simeq1.5\times1.5$~mm$^2$, Fig.~\ref{fig:LED_mount}~(a)] is far bigger compared to the cross sectional area of 200~$\mu$m optical fiber. This explains $\sim 0.1-0.2$\% efficiency of the optical output. Note that the use of 1~mm diameter optical fiber results in more than 20 times higher optical power. }

\end{itemize}

\begin{table}[htb]
\centering
\caption{\label{tab1} Characteristics of studied LEDs.  The meaning of the parameters are:  $\lambda_{\rm sp}$ is the wavelength as announced in LEDs specification sheet, $\lambda_{\rm av}$ is the average wavelength detected in experiments, FWHM is the full width at the half maximum, $P_{\rm opt}(0.7$~A) is the output optical power at the end of ${\o}=200$~$\mu$m optical fiber (NA=0.29) at applied current $J=0.7$~A, and $P_{\rm opt}(0.7{\rm ~A})/P_{\rm el}(0.7$~A) is the optical efficiency.}
\vspace{0.5cm}
\begin{tabular}{ccccccccccc}
\hline
\hline
$\lambda_{\rm sp}$&$\lambda_{\rm av}$&FWHM&$P_{\rm opt}(0.7$~A)&$P_{\rm opt}(0.7{\rm ~A})/P_{\rm el}(0.7$~A)\\
(nm)&(nm)&(nm)&(mW)& (\%)\\
\hline
395&402.8&21.6&4.21&0.18\\
430&426.0&11.6&4.00&0.16\\
470&471.6&21.6&1.77&0.78\\
520&523.9&24.9&1.92&0.84\\
530&531.9&34.2&1.49&0.66\\
600&597.0&94.6&3.96&0.18\\
730&731.4&31.5&--&--\\
\hline
\hline
\end{tabular}
\end{table}

\subsection{The fluorescence response of ruby and strontium tetraboite}\label{sec:ruby-SBO_responce}

\begin{figure*}[htb]
\centering
\includegraphics[width=0.9\linewidth]{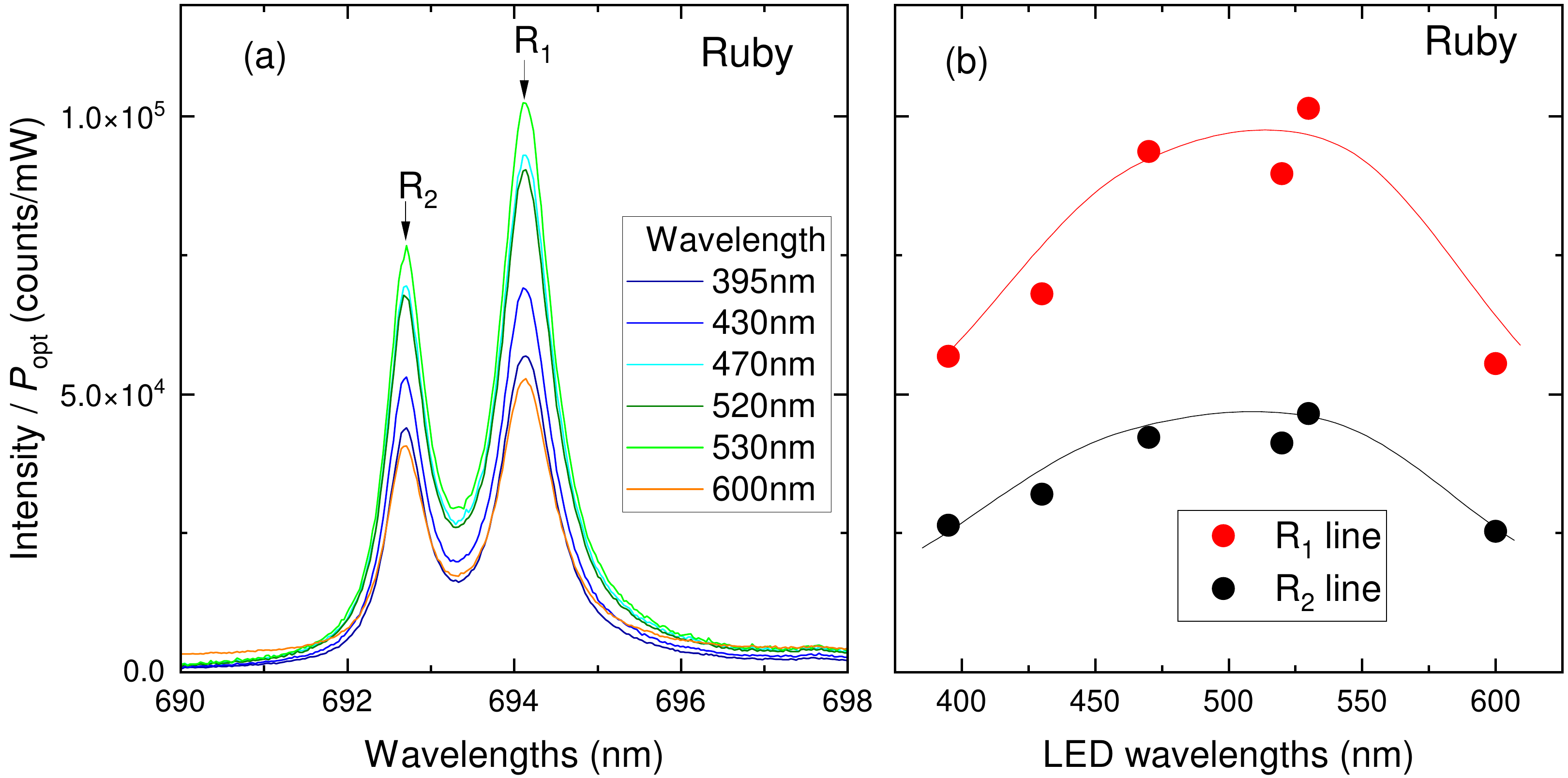}
\caption{(a) The fluorescence spectra of ruby (Cr$^{3+}$:Al$_2$O$_3$) at ambient pressure measured by using LEDs with the average wavelength ranging from 395~nm to 600~nm. $R_1$ and $R_2$ denote the corresponding fluorescence lines. (b) Dependence of the normalised intensity of $R_1$ and $R_2$ ruby lines on the LEDs wavelengths. The lines are guides for the eye.}
 \label{fig:fluorescence_ruby}
\end{figure*}

\begin{figure*}[htb]
\centering
\includegraphics[width=0.9\linewidth]{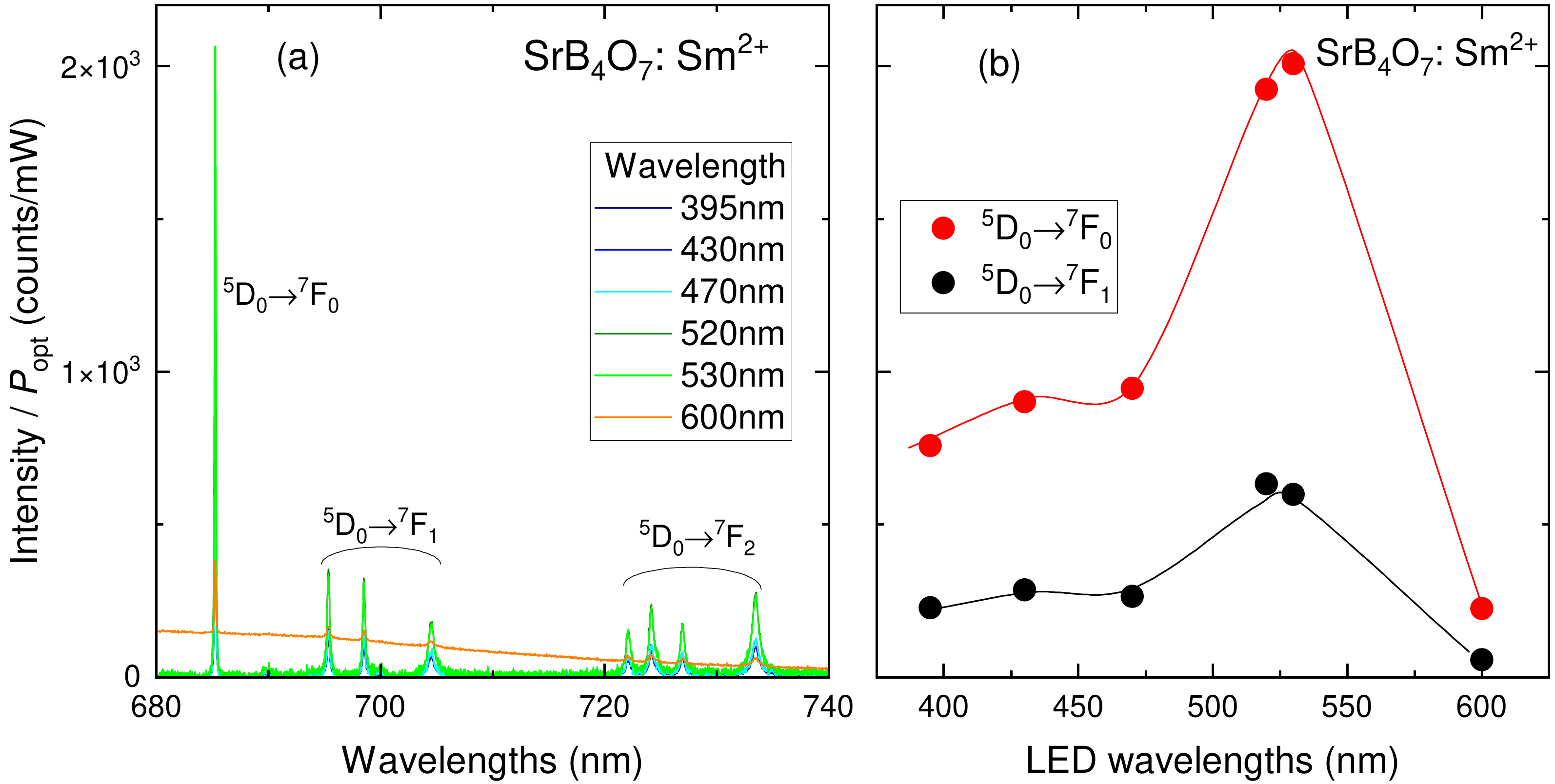}
\caption{(a) The fluorescence spectra of strontium tetraborate (Sm$^{2+}$:SrB$_4$O$_7$) at ambient pressure measured by using LEDs with the average wavelength ranging from 395~nm to 600~nm. Eight lines correspond to: $^5D_0 \rightarrow {^7F}_0$ (685.4 nm), $^5D_0 \rightarrow {^7F}_1$ (695.4, 698.6, and 704.6 nm), and $^5D_0 \rightarrow {^7F}_2$ (722.1, 724.2, 727.0 and 733.5 nm) and they are associated with the so-called  $4f^6-4f^6$ intra-configurational transitions of Sm$^{2+}$ ions.\cite{Zheng_SrB4O7_JMC_2020} (b) Dependence of the normalised intensity of  $^5D_0 \rightarrow {^7F}_0$ and the first  $^5D_0 \rightarrow {^7F}_1$ lines of strontium tetraborate on the LEDs wavelengths. The lines are guides for the eye.}
 \label{fig:fluorescence_SrBO}
\end{figure*}

Figures~\ref{fig:fluorescence_ruby} and \ref{fig:fluorescence_SrBO}  represent the results of fluorescence measurements of ruby (Cr$^{3+}$:Al$_2$O$_3$) and strontium tetraborate (Sm$^{2+}$:SrB$_4$O$_7$), respectively. Experiments were performed at ambient pressure by using fiber-coupled LED sources and an optical reflection probe. The pressure indicators were mounted inside especially designed holders allowing easy attachment to SMA905 optical connector.

The fluorescence spectra of ruby [Fig.~\ref{fig:fluorescence_ruby}~(a)] is confined in the region of $690 - 698$~nm and it consists of two lines ($R_1$ and $R_2$) which correspond to the transition of chromium from the excited metastable level $E_2$ to the ground state $E_1$.\cite{Mykhaylyk_ruby_Sensors_2020} The red emission spectra of Sm$^{2+}$ in SrB$_4$O$_7$ is located in the range of $680-740$~nm [Fig.~\ref{fig:fluorescence_SrBO}~(a)]. Eight lines correspond to: $^5D_0 \rightarrow {^7F}_0$ (685.4 nm), $^5D_0 \rightarrow {^7F}_1$ (695.4, 698.6, and 704.6 nm), and $^5D_0 \rightarrow {^7F}_2$ (722.1, 724.2, 727.0 and 733.5 nm) and they are associated with the so-called  $4f^6-4f^6$ intra-configurational transitions of Sm$^{2+}$ ions.\cite{Zheng_SrB4O7_JMC_2020}

The data presented in Figs.~\ref{fig:fluorescence_ruby}~(a) and \ref{fig:fluorescence_SrBO}~(a) imply that LEDs with the wavelengths ranging from 395 to 600~nm are suitable for observing the fluorescence response of both, the ruby and the strontium tetraborate, probes. There was no fluorescence signal detected for the 730~nm LED. In this case only the broad intensity distribution corresponding to the excitation spectra of the LED itself (similar to the one presented in Fig.~\ref{fig:LED_spectra}) was recorded.

The selection of the most efficient LED type could be made by comparing the dependence of the normalized intensity of fluorescence line(s) as a function of LED's average wavelengths. The corresponding dependencies for $R_1$ and $R_2$ lines of ruby, as well as for $^5D_0 \rightarrow {^7F}_0$ and the first  $^5D_0 \rightarrow {^7F}_1$ lines of strontium tetraborate are presented in Figs.~\ref{fig:fluorescence_ruby}~(b) and \ref{fig:fluorescence_SrBO}~(b), respectively. The best result is obtained for the green light LEDs with the average wavelengths $\sim 530$~nm.

\subsection{LED instead of laser light sources for pressure measurements}\label{sec:replacemement-of-laser-by-LED}

This section discusses the use of LEDs as light sources for pressure determination by means of fluorescence. Two types of pressure cells were tested, {\it i.e.} piston-cylinder and the diamond anvil cell (DAC).

\subsubsection{Piston-cylinder pressure cell}

Experiments were performed using the double-wall type piston-cylinder cell described in Refs.~\cite{Khasanov_HPR_2016, Shermadini_HPR_2017}. The optical pressure read was accessed via the 'double-volume` setup, where the space inside the cell is split between the 'sample` ($\simeq 0.3$~cm$^3$) and the 'optical` ($\simeq 1$~mm$^3$) volumes, respectively.\cite{Naumov_PRA_2022, Khasanov_JAP_2022}
Strontium tetraborate (Sm$^{2+}$:SrB$_4$O$_7$) was filled inside the 'optical` volume and played a role of a pressure indicator. The insertion of the excitation radiation and the read of the fluorescence responce were performed in way described in Fig.~1 of Ref.~\cite{Naumov_PRA_2022}.

The results obtained using different fiber coupled LEDs were similar to that reported in Fig.~\ref{fig:fluorescence_SrBO}.
Due to the relatively large amount of pressure gauge material possible in such a double-volume setup ($\sim 0.5-1.0$~mm$^3$, Ref.~\cite{Naumov_PRA_2022}), the intensive fluorescence response was easily detected down to  $10-20$\% of a maximum electric power supplied to the LED ($P_{\rm el}$$\sim 0.3-0.6$~W, $P_{\rm opt}\sim 0.5-1.0$~mW).

\subsubsection{Diamond anvil cell}

\begin{figure}[htb]
\centering
\includegraphics[width=0.9\linewidth]{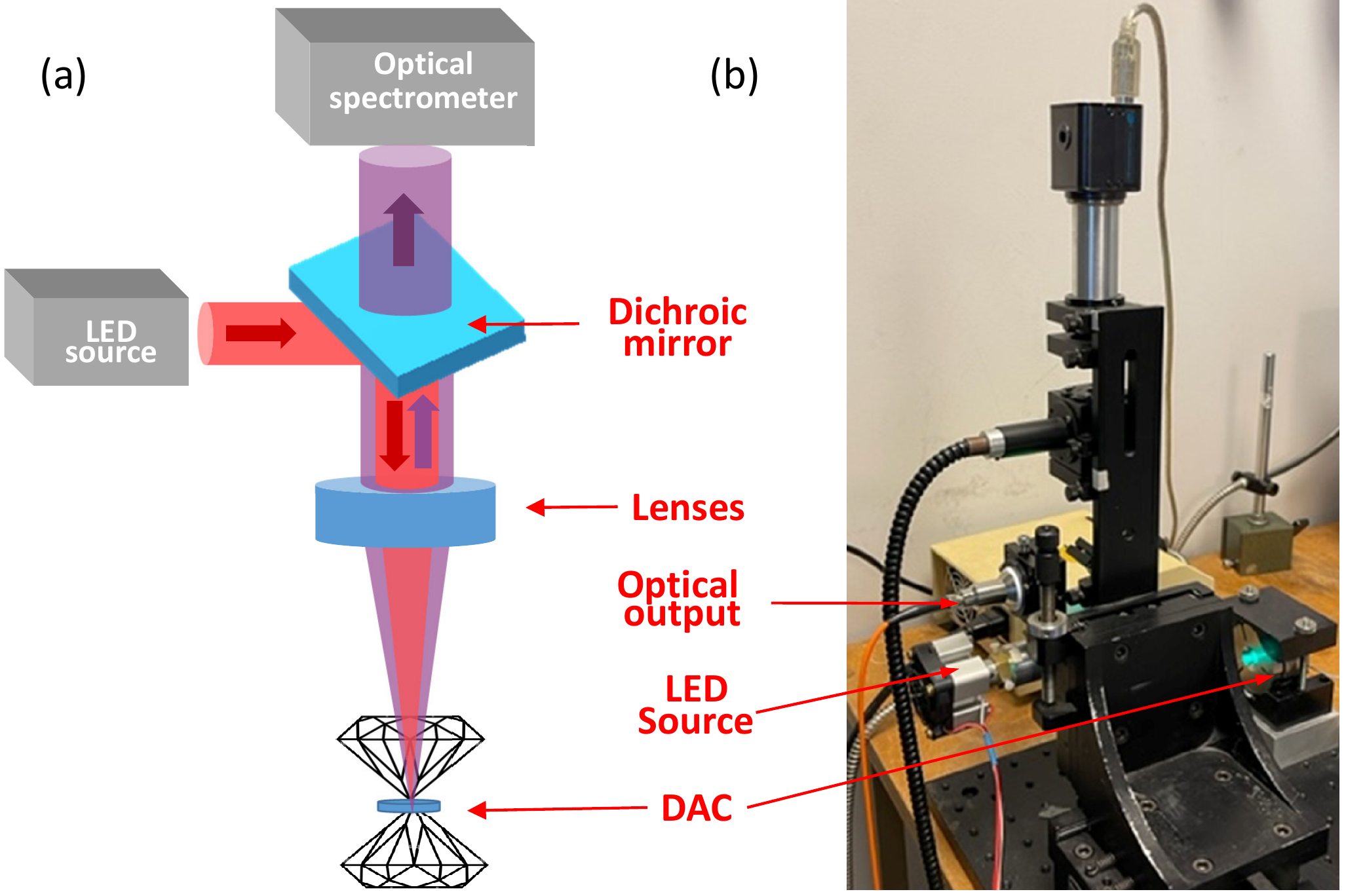}
\caption{(a) Schematic representation of DAC pressure measurement setup. The excitation light from the LED light source  is reflected by a dichroic mirror and transported through the system of lenses to the anvil cell. The fluorescence light, emitted by the pressure indicator placed in DAC, is transported back through the same lens system to the optical spectrometer.  (b) DAC pressure measurement setup at  the IMPMC laboratory (Sorbonne University, Paris). The lens holder of the LED source [Fig.~\ref{fig:LED_mount}~(c)] has the same outer diameter (${\o}=19$~mm) as the originally used 10~mW 532~nm green laser, which allows easy replacement of both light sources between the each other.}
 \label{fig:Laser-LED-setup}
\end{figure}

Experiments with a DAC were performed by using a standard fluorescence pressure measurement setup [see Fig.~\ref{fig:Laser-LED-setup}~(a) for schematics]. In such setup, the excitation light from the light source  is reflected by a dichroic mirror and transported through the system of lenses to the anvil cell. The fluorescence light, emitted by the pressure indicator (typically, a $\sim 10-20$~$\mu$m  ruby chip), is transported back through the same lens system to the optical spectrometer. The dichroic mirror plays the role of a filter by reflecting the intensive excitation light (the light from LED source in our case), but remaining transparent for fluorescence wavelengths. Compared to the simplified geometry presented in Fig.~\ref{fig:Laser-LED-setup}~(a), the real DAC setups are normally equipped with several additional elements as mirrors, objectives, filters, image cameras, fiber couplers {\it etc.} (see {\it e.g.} Refs.~\cite{betsa_spectrometer, dactools_spectrometer, almax_spectrometer}).

Figure~\ref{fig:Laser-LED-setup}~(b) shows the DAC pressure measurement setup at the IMPMC laboratory (Sorbonne University, Paris) with the LED source replacing the laser one. The lens holder of the LED source [Fig.~\ref{fig:LED_mount}~(c)] has the same outer diameter (${\o}=19$~mm) as the originally used 10~mW 532~nm green laser, which allows an easy exchange of two light sources between each other. The membrane DAC was loaded with ReO$_3$ powder sample together with a 4:1 methanol-ethanol mixture and a single $\simeq 15$~$\mu$m ruby sphere.  The cell was pressurised, initially, to $p\simeq6.5$~GPa.

\begin{figure}[htb]
\centering
\includegraphics[width=0.6\linewidth]{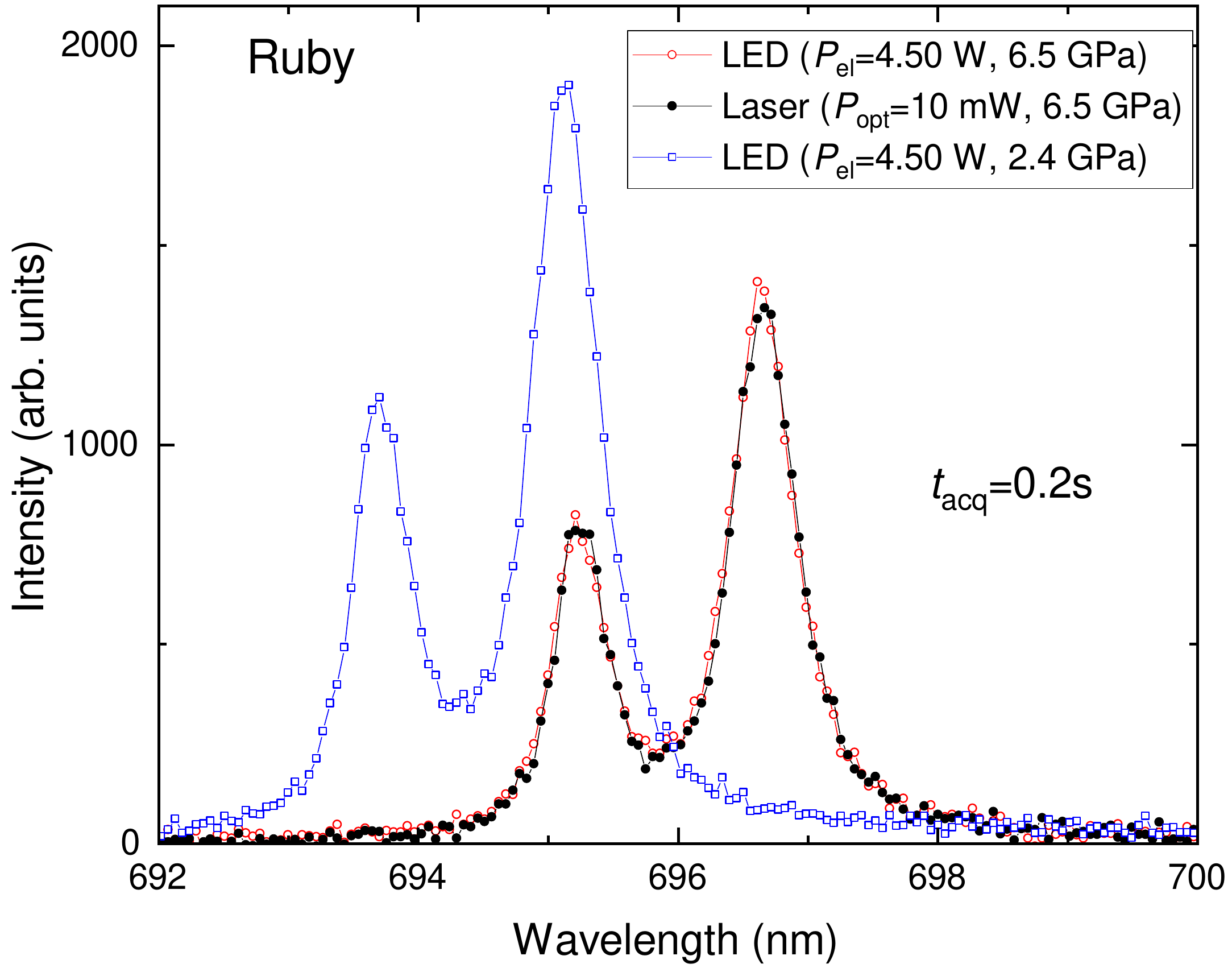}
\caption{Ruby fluorescence spectra collected in experiments with a DAC. The black and a the red curves correspond to the ruby florescence spectra at $p=6.5$~GPa as they are obtained using a laser (532~nm, $P_{\rm opt}=10$~mW) and a LED (520~nm $P_{\rm el}\simeq 4.5$~W, maximum electric power 10~W) light sources, respectively. The blue curve represent the response of ruby at $p=2.4$~GPa measured by using the 520~nm LED source.  }
 \label{fig:LED_ruby-responce}
\end{figure}

The results of the experiments by using the LED (520~nm, $P_{\rm el}\simeq 4.5$~W, maximum electric power 10~W) and the laser (532~nm, $P_{\rm opt}=$10~mW) light sources are presented in Fig.~\ref{fig:LED_ruby-responce}. The fluorescence spectra were accumulated during the same acquisition time $t_{\rm acq}=0.2$~s. Remarkably, the LED and the laser induced ruby fluorescence spectra are nearly undistinguishable from each other. It should also be noted, that results presented in Fig.~\ref{fig:LED_ruby-responce} were collected with the LED running at less than half of the maximum electric power ($P_{\rm el}\simeq 4.5$~W), meaning that for $P_{\rm el}$ exceeding 4.5~W,  the amplitudes of ruby fluorescence lines caused by LED source would exceed that of the laser.

As expected, by releasing the pressure, both ruby lines shift to lower wavelengths (blue curve in Fig.~\ref{fig:fluorescence_ruby}). The intensity of ruby lines increases by $\simeq 20$\%, which could be caused by slightly improved measuring conditions at lower pressures compared to the higher ones.  It is possible that the ruby chip had slightly moved and got less hidden by the ReO$_3$ powder sample.
%The probable reason is a slight increase of the gasket's inner hole allowing to collect more fluorescence light.

To check for reproducibility and for comparison with high- and low-power LEDs, similar measurements were performed for 520~nm and 530~nm LEDs with the maximum electric power limited to 3~W (Fig.~\ref{fig:LED_fiber} and Table~\ref{tab1}). For all three LED types and for the electric power up to $P_{\rm el}\sim 4.5$~W, the recorded ruby spectra stay close to each other.

\begin{figure}[htb]
\centering
\includegraphics[width=0.5\linewidth]{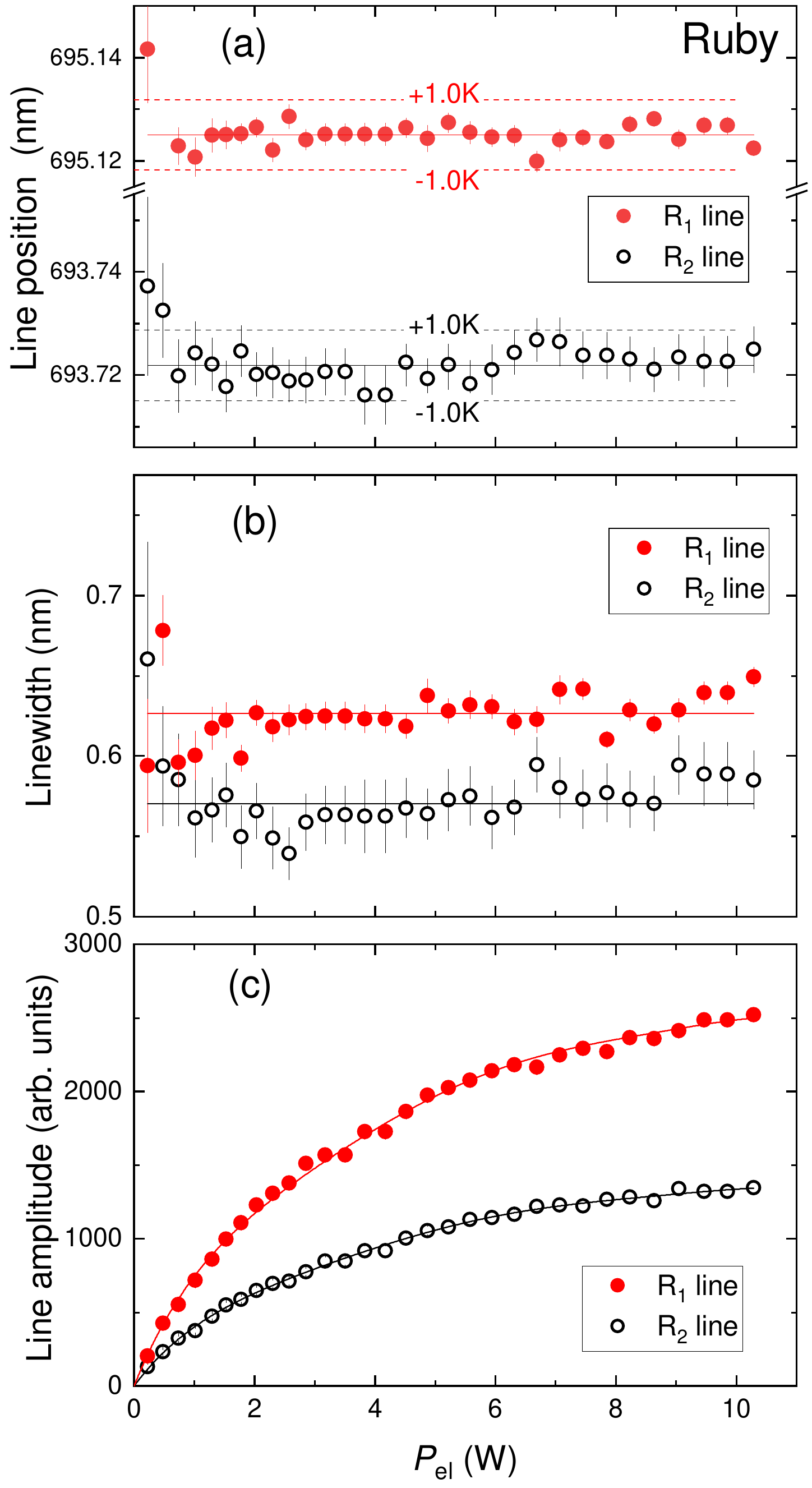}
\caption{(a) Dependence of $R_1$ and $R_2$ ruby line positions on the applied electric power $P_{\rm el}$ of 520~nm 10~W LED at $p\simeq 2.4$~GPa. The solid lines  are the zero-slope linear fits. The dashed lines represent the effect of temperature change by  $\pm1$~K (as obtained by following the relation ${\rm d} R_{1,2} /{\rm d}T = 0.0068$~nm/K, Refs.~\cite{Barnett_RSI_1973, Piermarini_RSI_1975}).  (b) Dependencies of $R_1$ and $R_2$ linewidths on $P_{\rm el}$. The solid lines  are the zero-slope linear fits. (c) Dependencies of $R_1$ and $R_2$ line amplitudes on $P_{\rm el}$.   The lines are polynomial fits.}
 \label{fig:LED_ruby_power-dependent-parameters}
\end{figure}

Figure~\ref{fig:LED_ruby_power-dependent-parameters} shows dependencies of several parameters of $R_1$ and $R_2$ ruby lines on the applied electric power to the 520~nm, 10~W LED measured at $p\simeq 2.4$~GPa. Fits were performed by using the Lorentzian $R_1$ and $R_2$ line shapes, which, for our case, lead to a better agreement compared to the Gaussian ones. It should be noted here that, in accordance with the literature data, the shape of both, $R_1$ and $R_2$, lines of ruby is neither Lorentzian, nor Gaussian. They are better described either by Voigt  or Moffat type of functions.\cite{Munro_JAP_1885, Shen_HPR_2021} Both these functional forms require, however, higher number of fitting parameters, as compared with the simple Gaussian or Lorentzian shapes. The simple Lorentzian fits, used in our study, may not describe the fluorescence spectral shape, but allow to capture dependencies of fit parameters on the applied electric power.

From the data presented in Figure~\ref{fig:LED_ruby_power-dependent-parameters} the following three important points emerge.
\begin{itemize}
  \item[(i)] $R_1$ and $R_2$ line positions [Fig.~\ref{fig:LED_ruby_power-dependent-parameters} (a)], as well as their linewidths [Fig.~\ref{fig:LED_ruby_power-dependent-parameters} (b)] stay independent of $P_{\rm el}$. Solid lines correspond to the zero-slope linear fits. In spite of much broader initial LED beam [$\simeq 1.5$~mm as is determined by the size of the LED chip, Fig.~\ref{fig:LED_mount}~(a)] compared with that of the laser source (typically 0.1~mm or less), such independence suggests that neither the ruby chip, nor the pressure cell are warmed up by the LED light.  For comparison, the $\pm1$~K dashed lines in Fig.~\ref{fig:LED_ruby_power-dependent-parameters}~(a) represent positions of ruby lines caused by the efect of temperature change by $\pm 1$~K (as obtained by following the relation ${\rm d} R_{1,2} /{\rm d}T = 0.0068$~nm/K, Refs.~\cite{Barnett_RSI_1973, Piermarini_RSI_1975}).
  \item[(ii)] The slope of $R_1$ and $R_2$ line amplitudes depend nonlinearly on the applied electric power [Fig.~\ref{fig:LED_ruby_power-dependent-parameters}~(c)]. This relates to the well known LED efficiency drop at high driving currents (see {\it e.g.} Ref.~\cite{Meyaard_APL_2013} and references therein). This effect is also clearly seen in Fig.~\ref{fig:LED_fiber}, representing dependence of the output optical power ($P_{\rm opt}$) on the applied electric power ($P_{\rm el}$) for different LED types.
  \item[(iii)] The efficiency drop means that with increasing $P_{\rm el}$ the fraction of the applied electric power wasted due to the heat should increase, while the fraction accessible for the light production would decrease in a proportional way. The data presented in Fig.~\ref{fig:LED_ruby_power-dependent-parameters}~(c) show that the doubling of the electric power ({\it i.e.} the increase of $P_{\rm el}$ from 5 to 10~W) results in only 20\% increase of $R_1$ and $R_2$ line amplitudes ({\it i.e.} in 20\% increase of the optical power). This implies that some LEDs (at least those types used in our studies) by being operated at half of the maximum electric power are capable to produce nearly 80\% of the maximum optical power. This may enhance the LEDs life time and simplify the heat sink construction.
 \end{itemize}
All together, the results presented in this section suggest that LEDs could successfully substitute laser sources in DAC pressure measurement setups. In most of the cases [as the one presented in Fig.~\ref{fig:Laser-LED-setup}~(b)] it might be just enough to replace the laser by the 'parallel light` LED source.

\section{Summary and Conclusions}\label{sec:Conclusions}

To summarise, this work discussed the use of light emitting diodes (LEDs) as a light source for pressure determination by means of fluorescence. A broad emitting spectra of commercial high-power LEDs do not prevent obtaining narrow fluorescence lines.  The best responses for two widely used type of pressure indicators, namely ruby (Cr$^{3+}$:Al$_2$O$_3$) and strontium tertaborite (Sm$^{2+}$:SrB$_4$O$_7$ ), were found for the green LEDs with the average wavelength $\lambda_{\rm av}\sim 530$~nm. The LEDs could be easily implemented for producing fiber coupled, as well as parallel light sources. Our experiments show that LED sources can successfully replace lasers in fluorescence pressure determination measurements for both the piston-cylinder and diamond anvil setups.

To conclude, the use of commercial LEDs  allows a simple and inexpensive construction of powerful light sources for fluorescence experiments. The incoherent light produced by LEDs is less dangerous for the human health and, as so, do not require as strong safety regulations as the coherent laser light. This permits the use of LED based pressure measurement setups at open platforms of large user-based facility instruments.

\section*{Data Availability Statement}

The data that support the findings of this study are available from the corresponding author upon reasonable request.

\section*{Acknowledgements}
The experiments were performed at the muon lab (LMU, PSI, Switzerland) and the Sorbonne University. The work was partially supported within the framework of 'ExtremeP` R'Equip project. RK acknowledges helpfull discussions at the LMU lab seminar. SK acknowledges assistance by K. B\'{e}neut and F. Datchi (IMPMC).

\end{document}